# Telling stories, making Hanzi: AI-assisted co-creation with elderly migrants in urban China

Yunfei Chen [a†], Wen Zhan[b†], Peiyue Lin[*c†], Ziqun Hua[*a†], Ying Hu[a]

[a] School of Design, Hunan University, China, People's Republic of

[b] Royal College of Art, United Kingdom

[c] University of the Arts London, Central Saint Martins, United Kingdom

[†] These authors contributed equally as first authors.

*Corresponding author e-mail: peiyue.design@gmail.com; helena_huahua@163.com

**Abstract**: This paper explores how older migrants in urban China can record stories that everyday language and design often miss. We ran two co-creation workshops with 10 elders. Activities combined oral storytelling, facilitator-mediated AI assistance, and hand-making. Large language models proposed candidate glyphs through a facilitator. Participants crafted new *Hanzi* to hold their stories. The resulting characters served as memory anchors for later sharing and retelling. Our interpretive analysis shows heterogeneity and adaptive capacity among participants. Participants experienced AI as a creative initiator that lowered barriers to expression and making, especially for those with lower digital literacy. The work challenges homogenizing assumptions about older adults and the presumption of uniform capacities and needs. We contribute a workshop framework that positions AI as a backstage facilitator. We also offer insights on engaging older migrants as sources of community memory and situated cultural knowledge within inclusive urban systems.

## 1. Introduction

With the rapid urbanization in China, an increasing number of older adults are relocating to cities, particularly in metropolitan areas and economically developed regions (Dou & Liu, 2017; State Council Information Office of People's Republic of China, 2018). This migration is often driven by family-related factors, such as joining adult children, providing grandchild care (Liu, 2014; Shi et al., 2023), or seeking better healthcare and employment opportunities (Dou & Liu, 2017). As a result, a mobile elderly population is emerging, embedded in complex intergenerational communities (Luo et al., 2012). These older migrants carry valuable memories of personal migration journeys, generational experiences, inter-city transitions, and societal changes (Dryjanska, 2015). However, these personal stories remain underrepresented. They are usually oral and fragmented, often overlooked by official written city records. While some older adults actively adopt digital tools, generational gaps in digital literacy still make it harder for many to engage with contemporary digital systems, constraining specific forms of civic and cultural participation (Kebede et al., 2022; Liam & Anne, 2015).

The exclusion of these narratives is not merely a matter of access but one of power. The inscription of narratives reflect control over which experiences are legitimised, preserved, and incorporated into mainstream narratives, with written language serving as a power medium in this process (Bourdieu, 1991; Foucault, 1972). In the Chinese context, the official written language, namely *Hanzi,* has long played a unifying role, which can inadvertently marginalize narratives conveyed through non-standard language, such as regional dialects or informal oral histories (Curzan et al., 2023; Lu et al., 2019). In this paper, we use *Hanzi* to refer to Chinese characters, a logographic writing system whose visual and compositional structure differs fundamentally from alphabetic phonetic scripts such as English. Here, *Hanzi*-making refers not to producing standardized new characters, but to using the compositional logic of Chinese characters as a symbolic storytelling practice. To challenge this narrative exclusion, some marginalised communities have historically engaged in practices that decentre official script and facilitate the re-creation of symbols as a form of cultural and political agency. For example, the Hmong people 苗族 in China also resisted the adoption of official writing systems to maintain the informality and fluidity of their language (Michaud, 2020). Women in Jiangyong, Hunan also created *Nüshu* 女书, a women's writing system historically used to phonetically represent local dialect speech, share personal experiences, and sustain bonds among women outside male-dominated literacy structures (Liu, 2004). Inspired by such practices, we suggest that symbolic co-creation may offer older migrants a tangible way to rearticulate personal narratives through shared symbolic forms.

While tangible narratives have proven vital for marginalised groups to articulate their identities and secure their histories (Abranches & Horton, 2023; Ankenbauer & Brewer, 2024), the rapid expansion of AI-assisted co-creation risks further exclusion. AI tools offer potential for narrative expression by lowering barriers to creation. Yet, existing works on AI co-creation predominantly target professionals, often presuming relatively high levels of AI and digital literacy and treating AI-generated content as final digital outcomes (Prasad et al., 2025; Winerö & Utterberg Modén, 2024). To address this gap in older migrant narratives, we conducted two co-creation workshops in Shenzhen, a city with a floating population of 14.32 million (Shenzhen Municipal Bureau of Statistics, 2021). Two workshops involved ten older migrants, combining autobiographical storytelling with an AI-assisted *Hanzi*-making process. We designed a low-intervention, human facilitator-mediated AI configuration in which a large language model operates backstage, retrieving related semantic characters as creation initiators. These characters are then adapted and reconstructed into tangible new *Hanzi* by older migrants, enabling them to express and inscribe their own stories. Against this backdrop, we examine the following research questions:

- RQ1: How can *Hanzi*-making serve as a medium for expressing unspoken or unwritable stories among older migrants?

- RQ2: How can a facilitator-mediated AI configuration support this co-design process with older migrants?

More broadly, this study asks how symbolic co-creation might help older migrants not only express difficult-to-write experiences, but also enter public cultural narratives from which they are often absent (Walsh et al., 2017). This paper contributes to design research in three interconnected ways. First, empirically, we show how *Hanzi*-making can function as a symbolic storytelling practice through which older migrants express personal and place-based experiences while being re-situated as cultural producers. Second, methodologically, we contribute a facilitator-mediated, low-intervention AI configuration that supports low-barrier co-creation without displacing participants' interpretive agency, thereby offering one possible approach to layered digital inclusion. Finally, we suggest that museums and other cultural venues can serve as mediating spaces for public cultural participation in later life, extending Design for Longevity toward questions of cultural visibility, expression, and belonging (Hodara, 2024; Lee et al., 2024).

## 2. Related work

### *2.1 From Tangible Narrative to Public Inclusion*

Tangible narrative practice is a key point of intervention in participatory design. Storytelling helps people reconstruct identity and meaning in changing situations (Bruner, 1990). In participatory design, co-making with materials helps marginalised groups externalize experiences and negotiate shared meanings in situ. Tangible media matter because they organise interaction and attention in embodied ways, supporting co-presence, coordination, and discussion (Hornecker & Buur, 2006). When such tangible, participatory narratives move into public space, materials act as conversational scaffolds that render personal memories visible and discussable (Askins & Pain, 2011). These tangible elements—such as objects or artworks—serve as mediators, providing physical anchors for memory, which enables participants to co-create meaning, validate personal histories, and foster a sense of belonging. They facilitate shared interpretation and action, for example, through participatory collage (Abranches & Horton, 2023; Mazzaro, 2022) or stop-motion animation with personal objects (Eivazy, Peeters, & Claes, 2023). By bringing these narratives into the public, this approach can challenge dominant narratives and build more inclusive community memories (Kambunga et al., 2020; Shefik, 2018). This aligns with age-friendly and placemaking goals that situate stories and activities in accessible, usable, and inclusive community and cultural venues (World Health Organization, 2007; Yuan et al., 2022).

Studies with older adults particularly underscore the value of this approach. Tangible objects and environments—such as keepsakes, memory objects, and interactive displays—play a crucial role in helping older adults recall memories and stimulate storytelling (Ankenbauer & Brewer, 2024). In specific contexts, such as Mild Cognitive Impairment (MCI), tailored methods like "photo narrative" show potential design space for supporting multidimensional expression and opportunities to support meaning-making (Aflatoony, Dorta, & Guité, 2020). Moreover, research has shown that converting narrative into empowering service actions contributes to strengthening agency and participation for older residents (Pan et al., 2019). Building on this literature, we situate our work in a community art museum and a community-based elder care institution as public sites, using materialized *Hanzi*-making to route individual expression into public participation, thereby strengthening visibility and social connection.

### *2.2 Challenges in AI-assisted Co-creation with Older Adults*

AI-assisted co-creation is a key topic in design and HCI (Fang et al., 2025). The work now extends from professional designers to workshops with non-designers. Studies show AI can support students' design-based learning (Prasad et al., 2025), and aid teachers in redesigning assessment (Winerö & Utterberg Modén, 2024). However, low AI literacy remains a barrier, as do structural power inequalities and tokenism (De Meulder et al., 2024; Farr, 2018).

At the same time, ageing and technology is a timely design agenda driven by demographics and policy priorities. Building on participatory design, prior work has proposed many collaborative pathways for workshops. Examples include direct involvement of older adults, the use of proxy participants, and methods such as probes, prototypes, and video prompts (Halskov & Hansen, 2015; Botero & Hyysalo, 2013; Lindsay et al., 2012). How participation is configured matters. Different arrangements enact different versions of aging, so participation should be organised with care (Fischer et al., 2021).

Two gaps persist at this intersection. First, a tooling gap: Mainstream creative tools often presume high digital fluency and offer little integration with tangible materials that many non-expert users rely on (Islam et al., 2023). Second, a homogenizing-assumption gap: mainstream AI tools tend to presume uniformly high literacies, while ageing design often presumes uniformly low literacies, overlooking differences that range from advanced AI users to non-users of smart devices.

To address the tooling gap, this study explores a "screen-free" AI interaction, combining LLM-based semantic suggestions with physical material manipulation. The heterogeneity gap and wider social challenges frame the key context for this exploration. This context suggests a focus not only on tools but also on how participation is organised (Fischer et al., 2021). We therefore use "translation work" (Botero et al., 2020) as a theoretical lens to reflexively analyze how "human facilitation" navigates the constraints and demands within this setting.

## 3. The workshop

We conducted two workshops (n = 10) at Pingshan Art Museum and a residential elder care facility in Pingshan, Shenzhen, respectively, to examine how older adults engaged in expressive storytelling and symbolic making. The sessions followed a practice-based participatory design approach in which narratives were developed through tangible handcrafting, while an AI-assisted co-creation tool provided optional prompts and structure. We built an interactive website (local prototype available at: https://figshare.com/s/5a1f6598421863b51cb7) to support idea capture and selection, with physical materials enabling the translation of stories into symbolic forms.

### *3.1 Sites and Participants*

In total, 10 older participants took part (see table 1). All were migrants to Shenzhen, and they spoke with regional dialect accents to varying degrees. Recruitment used the museum's community WeChat group and coordination with nursing home staff.
**First workshop (n = 3)**. The session was held in a museum affiliated community space. Eight people participated, including three elderly community members (mean age about 69), two museum staff, and three researchers (one facilitator and two note-takers). The museum outreach team and the authors co-planned and facilitated the session.
**Second workshop (n = 7).** The session was held in a residential elder care institution. Seventeen people participated, including seven residents (mean age about 84), two museum staff, two residential care staff, and four researchers (two facilitators and two note-takers). Care staff assisted with timing, consent flow, and safety procedures. Three older adults with disabilities initially joined the session: one with Alzheimer's-related cognitive impairment, one with a hearing impairment, and one wheelchair user with mobility and speech impairment. Due to fatigue, they chose to leave after the icebreaking activities and were therefore not included in the final participant count.

*Table 1 Participant Overview.*

| ID | **Age range** | **Region of Origin** | **Previous Occupation** | **Years in Shenzhen** |
|---|---|---|---|---|
| P1 | 51-60 | Northeast China | Accountant | 20 years |
| P2 | 71-80 | Central China | Public official | 10 years |
| P3 | 71-80 | Central China | Teacher | 10 years |
| P4 | 81-90 | Northeast China | Teacher | 5 years |
| P5 | 81-90 | Northeast China | Teacher | 5 years |
| P6 | 81-90 | Central China | Teacher | 30 years |
| P7 | 81-90 | East China | Factory worker | 30 years |
| P8 | 71-80 | East China | Engineer | 15 years |
| P9 | 81-90 | Northeast China | Teacher | 10 years |
| P10 | 81-90 | East China | Manager | 22 years |

## 3.2 System, Materials, and Prompt

### 3.2.1 *Hanzi*-making as the representational base

We adopted *Hanzi*-making as the primary form of expression to help participants externalize stories through recognizable cultural symbols. We used *Xiaozhuan* 小篆 as the visual base because it preserves pictographic qualities while remaining legible to contemporary readers. The interactive system drew from a digitized corpus of one thousand *Xiaozhuan* characters taken from the *Thousand Character Classic* 千字文 in the style of *Deng Shiru* 邓石如 (1743–1803).

### 3.2.2 Backstage interaction and retrieval workflow

Each story was transcribed by facilitators. To stabilise inputs and support relevant retrieval, facilitators structured the narrative into five fixed elements: Character, Cause, Process, Arrival, and Emotional Change. This structured text was entered into a custom website powered by DeepSeek. DeepSeek was used here as a retrieval aid rather than an autonomous generator of final outputs. We recognise that LLM-based retrieval may reflect biases embedded in dominant language data and may privilege more standardised or conventional associations. Rather than being adopted directly, AI suggestions were introduced into the workshop as prompts and mediated in situ through facilitator-participant discussion. Participants could interpret, accept, modify, or reject these suggestions during the co-creation process. Both workshops therefore adopted an indirect interaction strategy. AI operated in the background and returned *Xiaozhuan* character cues as semantic prompts. The website displayed three retrieved candidates for each story segment. Figure 1 illustrates the front-end facilitator input and the back-end retrieval process, and contrasts this indirect, retrieval-based approach with typical real-time generative co-creation workshops.

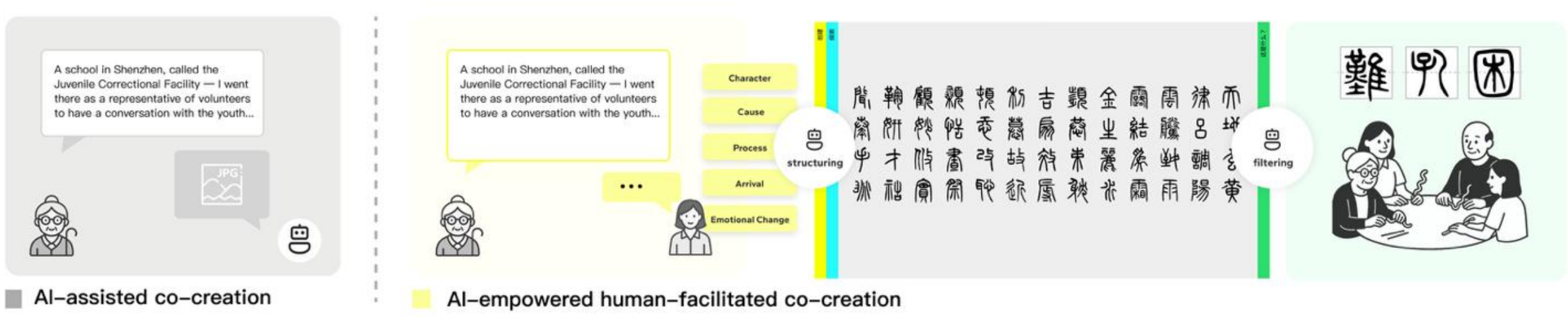

*Figure 1 Backstage AI-assisted workflow supporting narrative structuring and co-creation.*

### 3.2.3 Material testing and selection

Before the workshops, we ran a small material trial to evaluate which options older adults could handle within a short session (see figure 2). We compared ultra-light clay, pipe cleaners, air-dry clay, and foam board, considering comfort and clarity of form. Pipe cleaners and foam board provided the best balance of low physical effort and expressive shaping, so they were adopted in both sessions.

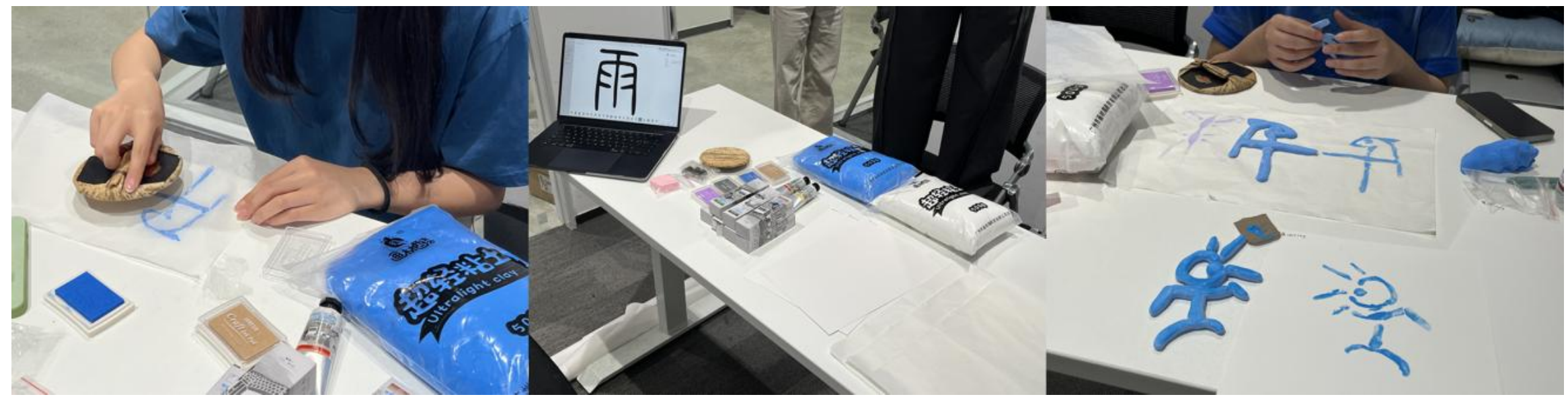

*Figure 2 Material testing evaluating options for older adults, testing ultra-light clay, air-dry clay, and foam board.*

## *3.3 Workshop Procedure*

This procedure aimed to keep narrative continuity with minimal screen interaction. Each two-hour workshop consisted of four main steps, as shown in Figure 3:

1. **Introduction and warm-up:** Team and participant introductions, followed by an overview of the workshop's goals and structure. A brief presentation introduced *Hanzi* formation principles, including the Six Scripts *(liùshū* 六书*)* and examples of invented *Hanzi* from dialects and online culture. Based on observations from the first workshop, we iterated this warm-up step for the second workshop by adding a collaborative icebreaker (see figure 4). Each small group made one part of a character with the provided materials. We then combined the parts to form a random composite character. The aim of this procedural adjustment was to lower the entry barrier, build joint attention, and activate tactile engagement.
2. **Storytelling & AI *Hanzi* Retrieval:** Participants shared personal stories about "travelling far from home." Facilitators transcribed the stories. As another iteration for the second workshop, facilitators used a set of prepared guiding questions (https://figshare.com/s/5a1f6598421863b51cb7) to help participants elaborate on their memories. The facilitators then structured the stories (as described in section 3.2) and entered them into the custom-built website. The system returned three relevant *Hanzi* in *Xiaozhuan,* which were printed and handed to participants.
3. ***Hanzi*-Making:** Using provided craft materials, participants created new *Hanzi* forms inspired by the retrieved *Hanzi*, combining physical composition with personal interpretation.
4. **Sharing and Archiving:** Participants named their pieces or explained meanings. We linked each piece to its narrative excerpt for archiving. In the nursing home group, staff helped arrange a brief collective display.

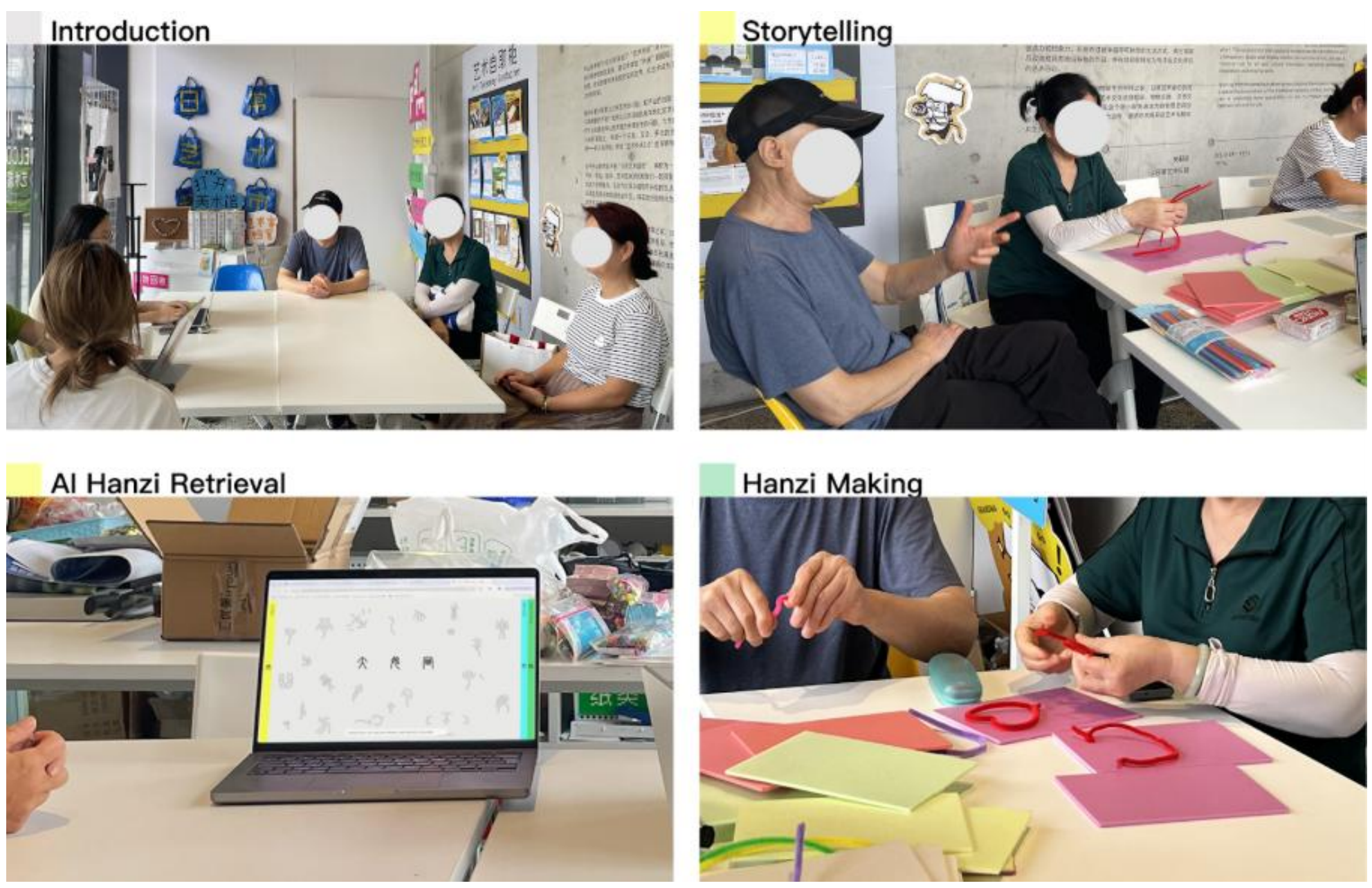

*Figure 3 Key stages of the workshop: introduction, story sharing, AI Hanzi retrieval, and Hanzi-making.*

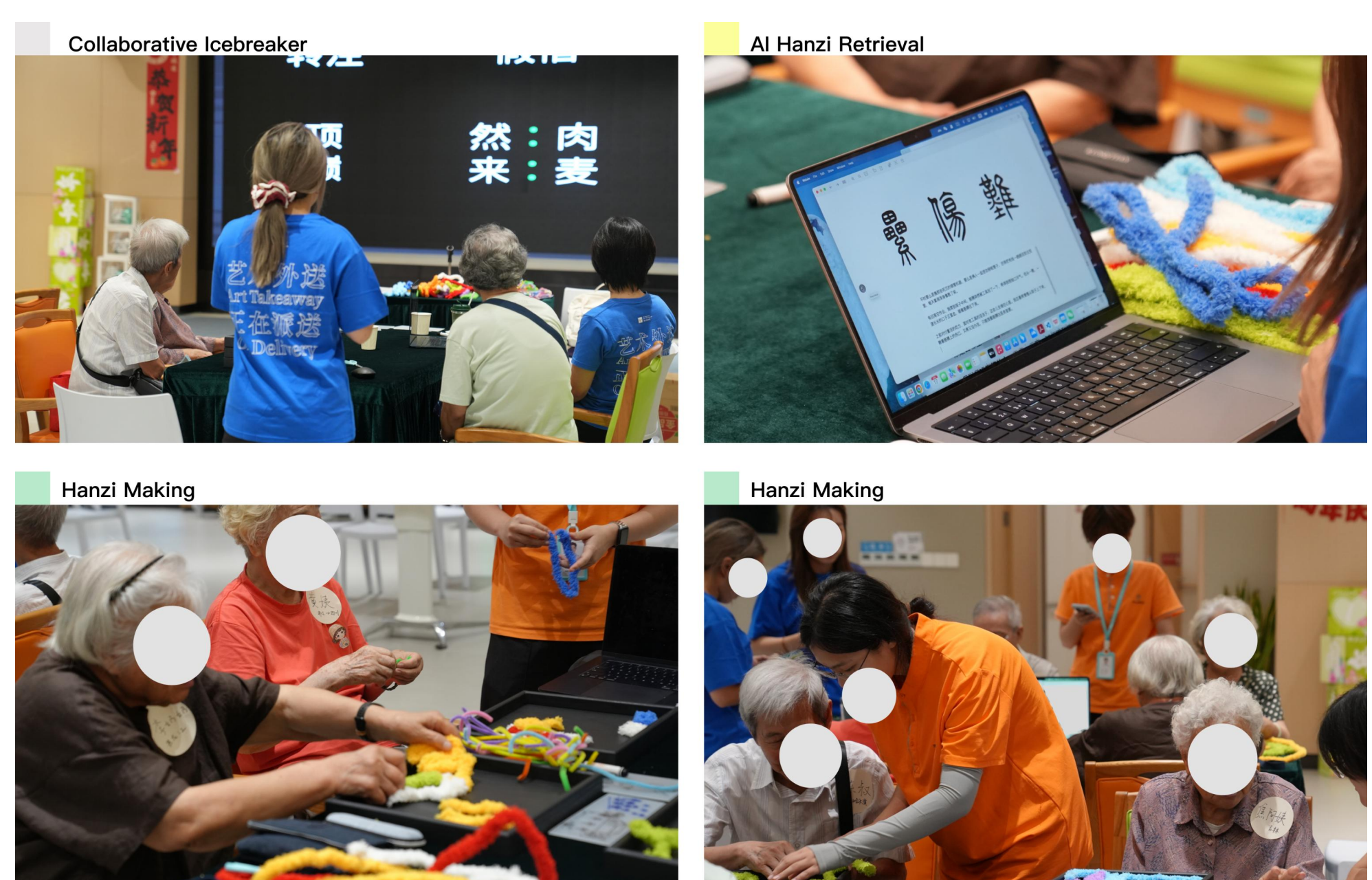

*Figure 4 Second workshop: warm-up, AI Hanzi retrieval, and Hanzi-making.*

## 3.4 *Data Collection and Analysis*

Data collection included documented stories, website input and output, photographs, participant-created artefacts, and field notes. Although audio recordings were initially planned, due to the presence of multiple participants, the noisy environment, and poor recording quality, we did not use it because it was difficult to match audio sources to specific participants. An interpretive analysis (Kadyschuk, 2023) was conducted to examine participants' verbal and material expressions.

### 3.5 *Ethics*

This research was approved by Hunan University Ethics Committee. All participants received an explanation of the study aims, workshop structure, and data handling procedures, and gave verbal consent for participation, recording, and documentation of the workshop process. Participation was voluntary, with the option to pause or withdraw at any time.

## 4. Findings

### 4.1 *Narrative Differentiation through Hanzi-Making*

In the two workshops, the process of *Hanzi*-making helped older participants transform previously undifferentiated "traveling far from home" experiences under the same theme into more personal and differentiated narratives. During this stage, participants were asked to select from their earlier storytelling the specific events, objects, or emotions that best represented their experiences and combine these elements into new character forms. This process required a cognitive reorganization of memory, translating temporal narration into spatial construction and giving abstract recollections a concrete visual shape.

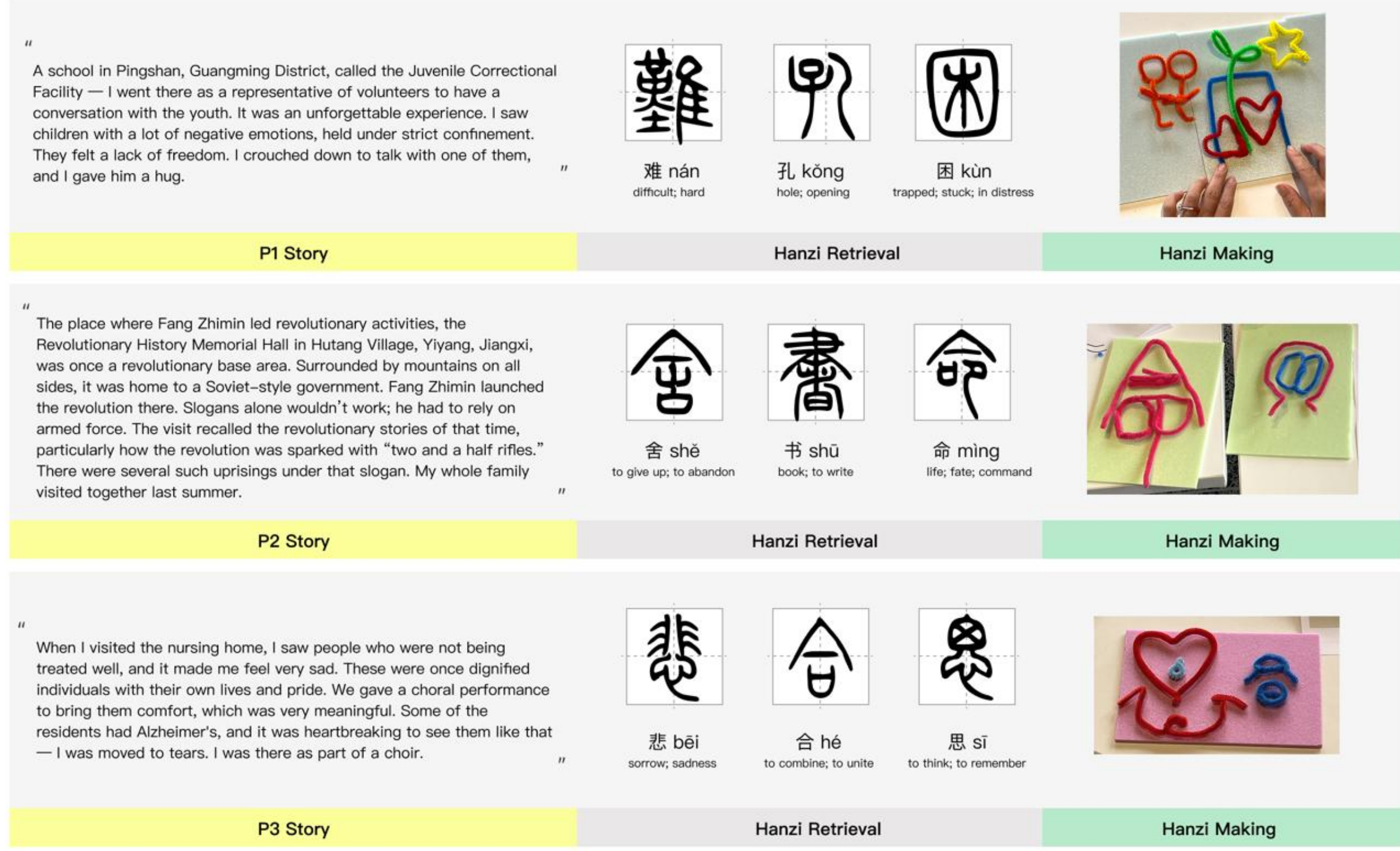

*Figure 5 Documentation of the first workshop participants' (P1, P2, and P3) narrative-based Hanzi creation process.*

Across the ten participants, three expressive orientations emerged through this selective transformation process (see figure 5 and 6). P4 and P5 emphasized bodily encounters with the landscape and the environment, turning details such as "the mountain is steep and requires climbing with both hands and feet" or "the ship is large, crowded, and surrounded by flowing water" into undulating, flowing, or clustered compositions. P6, P7, P8, and P10 represented traveling far from home as an experience of production, construction, or long-term labour. They tended to select components that signified work settings or technical operations, arranging them into vertically layered and continuous forms to evoke endurance and repetition. For instance, P7, recalling an injury during high-altitude installation, combined the components for "难" and "工" to express the intensity of physical effort. P1, P2, P3, and P9 interpreted the theme as an emotionally or ethically charged act of visiting, commemorating, or caring for others. P1, when describing a visit to a juvenile detention centre, incorporated both enclosed and open structures to symbolize "being confined yet still hopeful."

Through this process of re-creating narratives from spoken to visual form, participants not only gave tangible shape to their experiences but also produced artefacts that mediated collective understanding. When shared and discussed within the group, these new characters served as anchors for further reflection, allowing individual memories to become publicly legible and collectively reinterpreted. As a result, what initially appeared as similar accounts of "going away" unfolded into distinct stories shaped by different bodily, material, and emotional orientations.

### *4.2 Heterogeneity among Participants*

Both workshops revealed a high degree of heterogeneity among older participants. When AI was introduced, most participants described limited experience with smartphones or digital interfaces and had little prior exposure to AI. Yet in the second workshop, one participant mentioned attending an AI course, and another described using several AI tools to generate images and videos. These cases challenged the assumption that older adults are a uniformly "technologically disadvantaged" group, as participants with higher digital literacy expressed greater curiosity and expectations for deeper modes of interaction with AI.

Narrative themes also reflected generational differences. Participants from the museum group were generally younger, with an average age of about 70, and had lived through China's reform and urbanization period; their stories often revolved around mobility, entrepreneurship, and migration. Participants from the care institution were older, with an average age of about 83; their moves between cities were mostly shaped by early career assignments or the state allocation system, and their narratives tended to centre on work, family life, and nearby living environments (State Council Information Office of the PRC, 2018).

In the creative process of new *Hanzi*, participants demonstrated distinct interpretive strategies. The care institution group tended to follow pictographic or objective narrative logics, breaking down personal stories into concrete elements such as people, objects, and settings, and combining them in an orderly way. The museum group approached the task more symbolically, first abstracting their stories into meaningful scenes and then identifying corresponding visual representations. Age differences also brought variations in physical ability, which influenced material interaction. In the first museum workshop, participants were offered a range of materials, and most chose twistable sticks because of their familiarity and ease of use, often recalling experiences of making crafts with their grandchildren. In contrast, participants in the care institution found the same material less manageable, likely due to reduced hand dexterity or stamina. As described in the method section, the care institution group received more guided facilitation during warm-up activities to support engagement, yet their interaction style was not more fluid. They tended to value the outcome, preferring materials that could yield precise and refined results.

## *4.3 AI as Enabling Condition in the Creative Process*

At the outset, most participants expressed a lack of confidence in their ability to create new *Hanzi* from personal stories, citing concerns about their artistic or conceptual capabilities. However, upon learning that an AI system could offer *Hanzi* suggestions based on their narratives, they became noticeably more open to the creative task. While they did not fully understand the technical workings of the system, the idea that the AI could assist still helped ease their hesitation, and their expression softened from a frown to a more relaxed look.

When the AI presented candidate *Xiaozhuan* glyphs, many participants reacted with surprise and delight. They perceived a meaningful correspondence between their long narratives and the AI-generated characters, and several remarked that the system seemed to “understand” their stories. This perception fostered engagement, as participants discussed how well each character aligned with their stories. At this stage, beyond enhancing confidence, the AI also acted as an emotional motivator, stimulating curiosity and anticipation. Its outputs then served as creative initiators, offering anchors from which participants could begin their own symbolic interpretations and making processes.

Throughout the process, the AI remained in the background, with human facilitators transcribing and inputting participants’ stories. This indirect mode of interaction reduced pressure and allowed participants to focus on storytelling and making. As a quiet co-creator, the AI offered semantic prompts that supported expression while preserving participants’ interpretive agency. Its value, in this context, lay not in technical sophistication but in lowering the cognitive threshold of creative engagement.

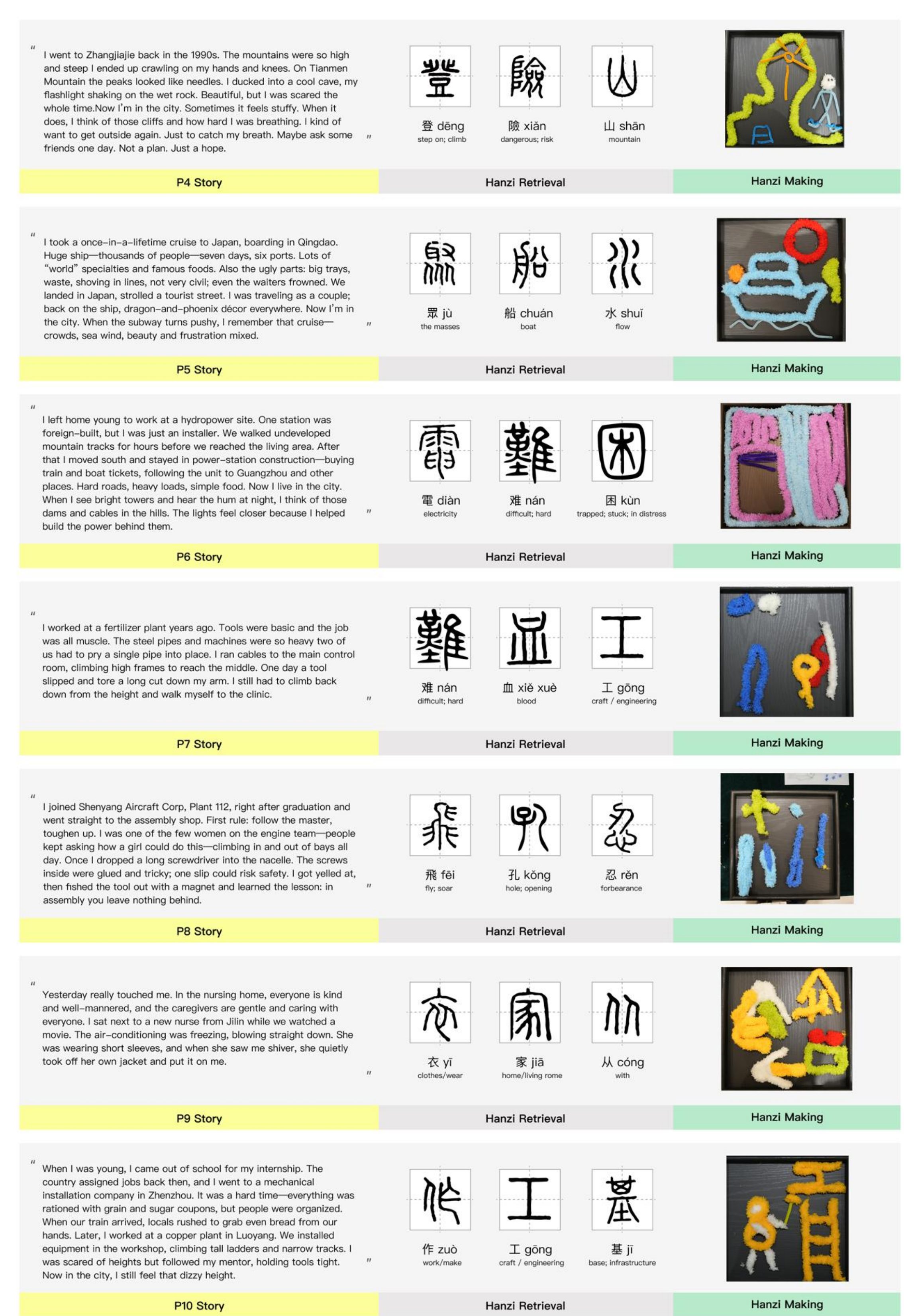

*Figure 6 Documentation of P4-P10's narrative-based Hanzi creation process.*

### 4.4 Observed role of the community museum

Before running the workshops, we conducted a short interview with staff at the community museum. They explained that regular visitors are mostly parents with children, and that older adults rarely come on their own. They linked this to older adults' limited experience with public cultural spaces and unfamiliarity with how to participate in them. Based on this, we set up the workshop with low digital demands and simple tangible materials, so that first-time participants could take part without extra technical preparation.

Using the same activity structure, the two sites nevertheless led to different forms of participation. In the community museum session (n = 3), the workshop acted as an entry point into local cultural programmes. After storytelling and *Hanzi*-making, participants gained a basic understanding of the museum and its public activities, and some asked whether similar events would be offered again. Here, the mediating role of the workshop was most visible: a single, low-cost making task created direct contact between older adults and a public cultural resource. In the elder-care institution session (n = 7), the same format mainly supported emotional connection and everyday engagement inside the institution. It provided a setting for residents to express themselves and be heard in a familiar environment. Although it generated fewer spontaneous links to wider community resources, it still introduced the museum as an available public channel.

This comparison suggests that the same co-creative format can be activated for different social functions across settings. In an open community museum, it more easily becomes a gateway for older adults to engage with public culture. In a more closed care environment, it is taken up as a tool for affective and routine participation, showing how local conditions shape the social meaning of design activities for older adults.

## 5. Discussion

### 5.1 Symbolic participation: from service recipients to cultural producers

The study demonstrated that the workshop effectively allowed participants to articulate memories that were rarely voiced in their everyday lives. Through Hanzi-making, they composed characters that embodied personal symbols—mountains, journeys, or gestures of care—transforming personal experiences into artefacts to share with others. This act of symbolic making enabled a form of authorship: it redefined language not as a system to be mastered but as a space of belonging to be inhabited through creative expression (Bruner, 1990; Kockelman, 2013). For participants, creating new *Hanzi* was less about reproducing correct forms than about inscribing their own experiences into a shared cultural code.

Such symbolic participation differs from extractive or instrumental forms of involvement often found in HCI (Sloane et al., 2022; Cooper et al., 2022). It is not about providing evaluative feedback but about reclaiming a voice in cultural systems. As Askins and Pain (2011) argue, the physical and embodied experiences of making in shared spaces can become a site of identity reconstruction and emotional connection. Our workshops illustrate this shift: older migrants moved from being designed for to designing with, acting as cultural producers whose stories and symbols enrich the collective narrative of the city. This repositions creativity as a form of civic agency embedded in making and storytelling within longevity-inclusive urban design (Le Dantec & DiSalvo, 2013; Lee & Sicklinger, 2024; Pan et al., 2019).

### *5.2 AI as Translation Infrastructure for Digital Inclusion*

Our facilitator-mediated configuration kept AI in the background to generate semantic prompts that were translated by facilitators into tangible cues. This reduced technical demands and positioned AI as a backstage collaborator rather than a primary interface. Although prior co-creation studies often involve direct AI use by designers or entrepreneurs (Kotturi et al., 2024; Guo et al., 2023), our approach relied on mediated interaction instead. This setup created a form of translation infrastructure (Botero et al., 2020), mediating between human capacities, materials, and algorithmic systems. It enabled participants to engage in narrative expression without requiring digital fluency, supporting an inclusive model in which AI amplifies expression through human mediation (Lindsay et al., 2012).
Within longevity-oriented systems, this suggests that facilitation is not an auxiliary aid but part of the digital infrastructure that configures how participation is organised (van Oorschot et al., 2022). By keeping AI backstage and routing interaction through human and material mediation, our configuration deliberately positioned older adults as the primary authors of the new characters rather than as data providers or recipients of AI-generated output. In this sense, lowering technical barriers was not only a matter of accessibility, but a way of deciding whose contributions would be recognised as central in an AI-mediated creative system. This AI-assisted process was not neutral. Because LLMs are shaped by dominant training data, some local and less publicly represented experiences may remain underrepresented (Gallegos et al., 2023). Facilitation therefore also served an interpretive role, helping participants to question, rework, or reject AI suggestions rather than treating them as authoritative outputs. While this did not remove algorithmic bias, it helped prevent AI outputs from being taken as definitive representations of participants' stories.

### *5.3 Heterogeneity as a Design Condition*

As shown in the findings, the two workshops involved older adults with varied educational backgrounds, AI familiarity, sensory abilities, and narrative orientations. This diversity reflects broader work that describes older adults as a heterogeneous population shaped by distinct life trajectories and socio-material conditions, rather than a single user type (Jarke, 2020; Kim, 2024). Recent research on conversational AI similarly argues that later-life technologies need to support different interaction modalities and configurations instead of one-size-fits-all solutions (Huang et al., 2025).
Our study captured only part of this heterogeneity. Prior work highlights further intersecting dimensions, including socioeconomic position, racialised histories, rural–urban differences, gender, and disability, all of which affect how older adults access, interpret, and benefit from digital systems (Marston & Samuels, 2019).

Acknowledging this broader field of variation helps situate our findings: the digital literacy differences we observed are one expression of a more complex landscape of ageing, technology, and inequality (Peine et al., 2015).

### *5.4 From Expression to Urban Literacy: Co-Authoring the City*

Our findings suggest that the value of narrative expression is not inherent in the act of creation itself but is closely linked to its social and spatial context. The contrast between the two sites highlights this distinction: while the closed context of the care institution tended to limit narratives to personal emotional connection, the museum's role as an open "mediating node" illustrated the potential for these personal stories to align with broader cultural infrastructures.

Seen through Design for Longevity and LongevityTech perspectives, this points to urban literacy as more than the ability to access services. It also involves writing oneself into the city by contributing memories and symbols to its shared cultural fabric (Hodara, 2024; Lee & Sicklinger, 2024). Community museums can be understood as cultural mediator nodes within urban service ecosystems (Huhn & Anderson, 2021). They host low-threshold activities, legitimise older adults' cultural contributions, and link personal expression to community assets and networks (Meroni & Selloni, 2022; Lee et al., 2024). Ultimately, this suggests a shift in longevity-oriented urban planning: cultural participation and symbolic authorship, supported by low-threshold infrastructures, should be treated as core components of age-inclusive systems, with cultural mediator nodes translating personal heritage into collective urban value.

## 6. Limitations

This study has several limitations. First, the two workshops involved a small and specific group: one session included only three older adults, and most participants across both sites were literate older adults who had lived in Shenzhen for an extended period. This limits the transferability of our findings to older participants with lower literacy, recent migration experience, less stable living conditions, or different access and support needs. Second, although some participants spoke with regional accents and had diverse life histories, we did not systematically include a broader range of intersecting differences, such as rural–urban divides, socioeconomic position, or disability. Third, the workshop design followed a largely linear structure that supported narrative continuity but offered limited opportunities for lateral interaction and peer-to-peer interpretation. Fourth, although facilitator mediation enabled participants to critically engage with AI suggestions during the workshop, potential algorithmic bias could not be fully mitigated. Taken together, these constraints suggest that our findings should be understood as context-specific and preliminary rather than universally generalisable.

## 7. Conclusion and Future Work

This paper introduced AI-assisted *Hanzi*-making to challenge narrative exclusion and inscribe the life experiences of older urban migrants. Through 2 co-creation workshops engaging 10 older migrants in Shenzhen, we showed how this low-intervention approach can function as a translation infrastructure, bridging the tooling and heterogeneity gaps in co-creation.

Our findings yield three main contributions. We first established that *Hanzi*-making is a powerful, culturally resonant medium for achieving narrative differentiation and assuming symbolic authorship. This mechanism enables participants to claim a form of civic agency by visually re-composing the official script to house their marginalised histories. Second, the facilitator-mediated AI configuration proved crucial for navigating participant heterogeneity. By positioning the AI backstage, we ensured that the technology amplified expression rather than automating or replacing human interpretive agency. Finally, systemically, we advance the agenda of Design for Longevity by redefining age-inclusive systems through the lens of urban literacy. The contextual difference observed between the museum and the care facility highlights the crucial role of spatial infrastructure. We argue that cultural participation must be treated as a core systemic component, requiring the integration of venues as cultural mediator nodes to translate personal heritage into collective urban value.

Future work will broaden participation to include older adults with lower literacy, stronger dialect use, shorter migration experience, or less stable living conditions, and experiment with alternative modes of AI involvement and facilitation to see how participation and narrative agency are configured in more varied settings. We also plan to collaborate with the partner museum and community organisations to integrate the co-created *Hanzi* and stories into cultural archives, educational activities, and programming, testing how these “tangible writings” might persist as part of the city’s collective memory and support ongoing age-inclusive cultural participation.

**Acknowledgements:** The authors would like to thank Lara Défayes and Romain Collaud for their valuable input in the design of the interactive website, and Flora Jiahui Ma for her support throughout the project. We are also deeply grateful to all the participants, as well as the staff at Pingshan Art Museum in Shenzhen, China, for their generous collaboration and assistance with this study. This work was supported by the Key Project of the Ministry of Education for Philosophy and Social Sciences of China. ChatGPT was used solely for language editing to improve clarity and readability. The authors take full responsibility for all content.

# 8. References

Abranches, M., & Horton, E. (2023). Heritage through collage: A participatory and creative approach to heritage making. *International Journal of Heritage Studies, 30*(1), 81–102. https://doi.org/10.1080/13527258.2023.2277780

Aflatoony, L., Dorta, T., & Guité, M. (2020). Photo-narrative for co-design with people with mild cognitive impairment. *Proceedings of the 16th Participatory Design Conference 2020*, 2, 111–115. ACM. https://doi.org/10.1145/3384772.3385143

Ankenbauer, S. A., & Brewer, R. N. (2024). Older adults’ personal curation of memory artefacts. *Proceedings of the ACM on Human-Computer Interaction, 8*(CSCW2), Article 456. ACM. https://doi.org/10.1145/3686995

Askins, K., & Pain, R. (2011). Contact zones: Participation, materiality and the messiness of interaction. *Environment and Planning D: Society and Space, 29*(5), 803–821. https://doi.org/10.1068/d11109

Botero, A., & Hyysalo, S. (2013). Ageing together: Steps towards evolutionary co-design in everyday practices. *CoDesign, 9*(1), 37–54. https://doi.org/10.1080/15710882.2012.760608

Botero, A., Hyysalo, S., Kohtala, C., & Whalen, J. (2020). Getting participatory design done: From methods and choices to translation work across constituent domains. *International Journal of Design, 14*(2), 17-34.

Bourdieu, P. (1991). Language and Symbolic Power. Harvard University Press.

Bruner, J. (1990). Acts of meaning. Harvard University Press.

Cooper, N., Horne, T., Hayes, G. R., Heldreth, C., Lahav, M., Holbrook, J., & Wilcox, L. (2022, April). A systematic review and thematic analysis of community-collaborative approaches to computing research. In *Proceedings of the 2022 CHI conference on human factors in computing systems* (pp. 1-18).

Curzan, A., Queen, R. M., VanEyk, K., & Weissler, R. E. (2023). Language Standardization & Linguistic Subordination. *Daedalus*, *152*(3), 18–35. https://doi.org/10.1162/daed_a_02015

De Meulder, M., Van Landuyt, D., & Omardeen, R. (2024). Lessons in co-creation: The inconvenient truths of inclusive sign language technology development. *arXiv preprint.* https://arxiv.org/abs/2408.13171

Deng, S. (1743–1803). *Thousand Character Classic* [Calligraphy in the style of Deng Shiru].

Dou, X., & Liu, Y. (2017). Elderly Migration in China: Types, Patterns, and Determinants. *Journal of Applied Gerontology*, *36*(6), 751–771. https://doi.org/10.1177/0733464815587966

Dryjanska, L. (2015). A social psychological approach to cultural heritage: Memories of the elderly inhabitants of Rome. *Journal of Heritage Tourism*, 10(1), 38–56. https://doi.org/10.1080/1743873X.2014.940960

Eivazy, N., Peeters, J. P. L., & Claes, S. (2023). Connecting through objects: Sharing memories through participatory stop-motion animation with personal objects of the Armenian diaspora. *Proceedings of the 11th International Conference on Communities & Technologies*, 48–53. ACM. https://doi.org/10.1145/3593743.3593756

Fang, C., Zhu, Y., Fang, L., Long, Y., Lin, H., Cong, Y., & Wang, S. (2025). Generative AI-enhanced human-AI collaborative conceptual design: A systematic literature review. *Design Studies, 97*, 101300. https://doi.org/10.1016/j.destud.2025.101300

Farr, M. (2018). Power dynamics and collaborative mechanisms in co-production and co-design processes. *Critical Social Policy, 38*, 623 - 644. https://doi.org/10.1177/0261018317747444

Fischer, B., Östlund, B., & Peine, A. (2021). Design multiple: How different configurations of participation matter in design practice. *Design Studies, 74*, 101016. https://doi.org/10.1016/j.destud.2021.101016

Foucault, M. (1972). *The Archaeology of Knowledge: And the Discourse on Language*. Knopf Doubleday Publishing Group.

Gallegos, I., Rossi, R., Barrow, J., Tanjim, M., Kim, S., Dernoncourt, F., Yu, T., Zhang, R., & Ahmed, N. (2023). Bias and Fairness in Large Language Models: A Survey. *Computational Linguistics, 50, 1097-1179*. https://doi.org/10.1162/coli_a_00524

Guo, X., Xiao, Y., Wang, J., and Ji, T. (2023). Rethinking designer agency: A case study of co-creation between designers and AI. *Proceedings in IASDR 2023: Life-Changing Design*. DRS. https://doi.org/10.21606/iasdr.2023.478

Halskov, K., & Hansen, N. B. (2015). The diversity of participatory design research practice at PDC 2002–2012. *International Journal of Human-Computer Studies, 74*, 81–92. https://doi.org/10.1016/j.ijhcs.2014.09.003

Hodara, E. S. (2024). LongevityTech: Bridging immersive media and design for longevity. *diid, 1*(82). https://doi.org/10.30682/diid8224f

Hornecker, E., & Buur, J. (2006). Getting a grip on tangible interaction: A framework on physical space and social interaction. *Proceedings of the SIGCHI Conference on Human Factors in Computing* Systems, 437–446. ACM. https://doi.org/10.1145/1124772.1124838

Huang, Y., Zhou, Q., & Piper, A. M. (2025, April). Designing conversational AI for aging: A systematic review of older adults' perceptions and needs. In *Proceedings of the 2025 CHI Conference on Human Factors in Computing Systems* (pp. 1-20).

Huhn, A., & Anderson, A. (2021). Promoting Social Justice through Storytelling in Museums. *Museum & Society, 19*(3), 351-368. https://doi.org/10.29311/mas.v19i3.3775

Islam, M. N., Khan, N. I., Inan, T. T., & Sarker, I. H. (2023). Designing User Interfaces for Illiterate and Semi-Literate Users: A Systematic Review and Future Research Agenda. *Sage Open, 13*(2). https://doi.org/10.1177/21582440231172741

Jarke, J. (2020). Ageing Societies and technological innovation. In Co-creating Digital Public Services for an Ageing Society. *Public Administration and Information Technology, 6*, 5-13. Springer. https://doi.org/10.1007/978-3-030-52873-7_2

Kadyschuk, L. (2023). Interpretive analysis. In J. M. Okoko, K. D. Walker, & S. Tunison (Eds.), *Varieties of Qualitative Research Methods: Selected Contextual Perspectives,* 249–255. Springer.

Kambunga, A. P., Winschiers-Theophilus, H., & Smith, R. C. (2020). Participatory memory making: Creating postcolonial dialogic engagements with Namibian youth. *Proceedings of the 2020 ACM Conference on Designing Interactive Systems*, 1069–1081. ACM. https://doi.org/10.1145/3357236.3395441

Kebede, A. S., Ozolins, L.-L., Holst, H., & Galvin, K. (2022). Digital Engagement of Older Adults: Scoping Review. *Journal of Medical Internet Research*, *24*(12), e40192. https://doi.org/10.2196/40192

Kim, M. (2024). Exploring autonomy as a design principle: Theoretical review of autonomy and case studies of service design for seniors. *Design and Culture, 16*(1), 83–107. https://doi.org/10.1080/17547075.2023.2259658

Kockelman, P. (2013). *Agent, Person, Subject, Self: A Theory of Ontology, Interaction, and Infrastructure*. Oxford University Press.

Kotturi, Y., Anderson, A., Ford, G., Skirpan, M., & Bigham, J. P. (2024, May). Deconstructing the veneer of simplicity: Co-designing introductory generative AI workshops with local entrepreneurs. In *Proceedings of the 2024 CHI Conference on Human Factors in Computing Systems,* 1-16.

Lee, S.-H., & Sicklinger, A. (2024). Design for longevity: People, process, and platform. *diid, 1*(82). https://doi.org/10.30682/diid8224a

Lee, S., Coughlin, J. F., Hodara, S., Meroni, A., & Sedini, C. (2024). Design for longevity (D4L): Project your future self through service and technology. *Proceedings of Design Research Society Conference 2024*. DRS. https://doi.org/10.21606/drs.2024.102

Le Dantec, C. A., & DiSalvo, C. (2013). Infrastructuring and the formation of publics in participatory design. *Social Studies of Science, 43*(2), 241–264. http://www.jstor.org/stable/432841811.

Liam, F., & Anne, B. (2015). *In Defence of Welfare 2*. Policy Press.

Lindsay, S., Jackson, D., Schofield, G., & Olivier, P. (2012). Engaging older people using participatory design. In CHI '12: *Proceedings of the SIGCHI Conference on Human Factors in Computing Systems*, 1199–1208. ACM. https://doi.org/10.1145/2207676.2208570

Liu, F. (2004). From being to becoming: Nüshu and sentiments in a Chinese rural community. *American Ethnologist*, 31(3), 422–439. https://doi.org/10.1525/ae.2004.31.3.422

Liu, J. (2014). Ageing, migration and familial support in rural China. *Geoforum*, *51*, 305–312. https://doi.org/10.1016/j.geoforum.2013.04.013

Lu, S., Chen, S., & Wang, P. (2019). Language barriers and health status of elderly migrants: Micro-evidence from China 老年流动人口的语言障碍与健康状况——来自中国的微观证据. China Economic Review, 54, 94–112. https://doi.org/10.1016/j.chieco.2018.10.011

Luo, Y., LaPierre, T. A., Hughes, M. E., & Waite, L. J. (2012). Grandparents Providing Care to Grandchildren: A Population-Based Study of Continuity and Change. *Journal of Family Issues*, *33*(9), 1143–1167. https://doi.org/10.1177/0192513X12438685

Marston, H. R., & Samuels, J. (2019, March). A review of age-friendly virtual assistive technologies and their effect on daily living for carers and dependent adults. In *Healthcare, 7*(1), 49. MDPI.

Mazzaro, C. (2022). The power of collage as a cultural heritage technique. *International Journal of Heritage Studies, 28*(11), 1193–1213. https://doi.org/10.1080/13527258.2022.2096530

Meroni, A., & Selloni, D. (2022). Service Design for Urban Commons. Springer.

Michaud, J. (2020). The Art of Not Being Scripted So Much. *Current Anthropology*. https://doi.org/10.1086/708143

Pan, S., Sarantou, M., & Miettinen, S. (2019). Design for care: How the "good old days" can empower senior residents to achieve better services in an aged-care institution. *The International Journal of Design in Society, 13*(2), 25–40. https://doi.org/10.18848/2325-1328/CGP/v13i02/25-40

Peine, A., Faulkner, A., Jäger, A., & Moors, E. (2015). Science, technology and ageing: Moving beyond the binary of innovation and need. *Technological Forecasting and Social Change, 93*, 54–67. https://doi.org/10.1016/j.techfore.2014.02.014

Prasad, P., Balse, R., & Balchandani, D. (2025). Exploring multimodal generative AI for education through co-design workshops with students. *Proceedings of the ACM Conference on Human Factors in Computing Systems.* ACM. https://doi.org/10.1145/3706598.3714146

Shefik, A. (2018). Reimagining transitional justice through participatory art. *International Journal of Transitional Justice, 12*(2), 314–333. https://doi.org/10.1093/ijtj/ijy013

Shenzhen Municipal Bureau of Statistics. (2021, May 17). The Seventh National Population Census Communiqué [No. 1] — Permanent resident population of Shenzhen. https://tjj.sz.gov.cn/zwgk/zfxxgkml/tjsj/tjgb/content/post_8771927.html

Shi, T., Hu, T., & Fan, X. (2023). Social capital and mental wellbeing of older people migrating along with adult children in Shenzhen, China. *International Journal of Environmental Research and Public Health, 20*(19), 6857.

Sloane, M., Moss, E., Awomolo, O., & Forlano, L. (2022, October). Participation is not a design fix for machine learning. In *Proceedings of the 2nd ACM Conference on Equity and Access in Algorithms, Mechanisms, and Optimization* (pp. 1-6).

State Council Information Office of the People's Republic of China. (2018, December 12). Progress in human rights over the 40 years of reform and opening up in China. http://english.www.gov.cn/archive/white_paper/2018/12/13/content_281476431737638.htm

van Oorschot, R., Snelders, D., Kleinsmann, M., & Buur, J. (2022). Participation in design research. *Design Studies, 78*, 101073. https://doi.org/10.1016/j.destud.2021.101073

Pan, J., Lindsay, S., Pritchard, G., & Olivier, P. (2013). Configuring participation: On how we involve people in design. P*roceedings of the SIGCHI Conference on Human Factors in Computing Systems,* 429–438. ACM. https://doi.org/10.1145/2470654.2470716

Walsh, K., Scharf, T., & Keating, N. (2017). Social exclusion of older persons: A scoping review and conceptual framework. *European Journal of Ageing*, *14*(1), 81–98. https://doi.org/10.1007/s10433-016-0398-8

Winerö, E., & Utterberg Modén, M. (2024). Engaging teachers in co-designing examinations for secondary schools in the era of large language models. *Proceedings of the First International Workshop on Participatory Design & End-User Development - Building Bridges*. https://ceur-ws.org/Vol-3778/short2.pdf

World Health Organization. (2007). *Global Age-friendly Cities: A Guide.* WHO Press.

Yuan, S., Wang, T., Zhang, S., & Wu, X. (2022). Patterns of living lost? Measuring community participation and other influences on the health of older migrants in China. *International Journal of Environmental Research and Public Health, 19*(8), 4542.